\documentclass[12 pt]{article}

\usepackage[english]{babel}
\usepackage[utf8x]{inputenc}
\usepackage{natbib}
\usepackage{amsmath}
\usepackage{amssymb}
\usepackage{gensymb}
\usepackage{authblk}
\usepackage{bbm}
\usepackage{bm}
\usepackage{geometry}
\usepackage{graphicx}
\usepackage{animate}
\usepackage{media9}
\usepackage{mathtools}
\usepackage{mathrsfs}
\usepackage{subcaption}
\usepackage{mwe}
\usepackage{xcolor}
\usepackage{hyperref}
\usepackage{multirow}
\usepackage{adjustbox}
\usepackage{lipsum}
\usepackage{float}
\usepackage{titlesec}
\usepackage[colorinlistoftodos]{todonotes}
\newcommand{\twopartdef}[4]
{
	\left\{
		\begin{array}{ll}
			#1 & \mbox{} #2 \\
			#3 & \mbox{} #4
		\end{array}
	\right.
} 

\title{A Test for Isotropy on a Sphere using Spherical Harmonic Functions }
\author{Indranil Sahoo, Joseph Guinness and Brian J. Reich}
\affil{\small{\it{Department of Statistics, North Carolina State University}}}
\date{\vspace{-5ex}}
\begin{document}
\maketitle

\begin{abstract}
\noindent Analysis of geostatistical data is often based on the assumption that the spatial random field is isotropic. This assumption, if erroneous, can adversely affect model predictions and statistical inference. Nowadays many applications consider data over the entire globe and hence it is necessary to check the assumption of isotropy on a sphere. In this paper, a test for spatial isotropy on a sphere is proposed. The data are first projected onto the set of spherical harmonic functions. Under isotropy, the spherical harmonic coefficients are uncorrelated whereas they are correlated if the underlying fields are not isotropic. This motivates a test based on the sample correlation matrix of the spherical harmonic coefficients. In particular, we use the largest eigenvalue of the sample correlation matrix as the test statistic. Extensive simulations are conducted to assess the Type I errors of the test under different scenarios. We show how temporal correlation affects the test and provide a method for handling temporal correlation. We also gauge the power of the test as we move away from isotropy. The method is applied to the near-surface air temperature data which is part of the HadCM3 model output. Although we do not expect global temperature fields to be isotropic, we propose several anisotropic models with increasing complexity, each of which has an isotropic process as model component and we apply the test to the isotropic component in a sequence of such models as a method of determining how well the models capture the anisotropy in the fields.

\medskip

\noindent {\it Key words and phrases:}
Spatial statistics, Anisotropy, Spherical harmonic representation.
\end{abstract}

\newpage 

\section{Introduction}

Modeling spatial dependence is a major challenge when analyzing geostatistical data. It is common to assume that the spatial covariance function is isotropic, meaning that the correlation between observations at any two locations depends only on the distance between those locations and not on their relative orientation \citep{guan2004nonparametric}. With advancements in technology we now observe massive amounts of data, especially in atmospheric sciences. Satellites and ground-based monitoring stations collect data, and large-scale climatic models produce data covering the entire globe. Thus it is important to develop methods for analyzing spatial data observed on spheres. In order to do so, it is necessary to understand the inherent correlation structure of the process on the sphere. Assuming that the process is isotropic will lead to simpler interpretation of the correlation structure and reduce computational complexity. However, in many applications, isotropy may not be a reasonable assumption and will lead to erroneous model fitting and predictions. 

A common method for checking for isotropy is to compare sample semi-variograms for different directions \citep{cressie1985, cressie1980, cressie1993}. Many approaches consider a stationary alternative and use directional variograms to construct test statistics \citep{matheron1961precision,  diggle1981binary,  cabana1987affine, baczkowski_mardia1990test, isaaks2001introduction}. Some nonparametric methods for checking isotropy are based on estimates of the variogram or covariogram \citep{lu2001testing,guan2004nonparametric,maity2012testing}. The notion of testing for second-order properties using the asymptotic joint normality of sample variogram evaluated at different spatial lags was established by \cite{lu2001testing}. The subsequent works of \cite{guan2004nonparametric} and \cite{maity2012testing} are based on these ideas. \cite{li2007nonparametric,li2008asymptotic} and \cite{jun2012test} consider spatiotemporal data and use approaches similar to the methods from \cite{lu2001testing}, \cite{guan2004nonparametric} and \cite{maity2012testing}. \cite{bowman2013inference} give a more computational approach for testing isotropy in spatial data using a robust form of the empirical variogram based on a fourth-root transformation. 

\cite{haskard2007anisotropic} extends the Mat\'ern correlation to include anisotropy, which facilitates a test of isotropy. \cite{fuentes2007approximate} describes a method in the spectral domain which is based on the estimation of parameters governing the directionality in the spatial dependence (anisotropy) using approximate likelihoods. \cite{matsuda2009fourier} once again consider a generalized Mat\'ern class which allows for anisotropy and constructs a likelihood ratio test for isotropy. 

All the methods discussed above are for random fields on the Euclidean space, $\mathbb{R}^d, d > 1$ and have stationarity as the alternative. However, the sphere is different from the Euclidean space in the sense that there is no notion of stationarity or geometric anisotropy on the sphere. Thus one cannot apply any of the above-mentioned tests for checking if the covariance function of a process on the sphere is isotropic. This necessitates the development of a test which will allow us to test for isotropy on the sphere.

Our approach for testing for isotropy on a sphere is similar in spirit to \cite{bandyopadhyay2017test} where the authors propose a test for stationarity on Euclidean spaces based on the discrete Fourier transform (DFT) vector, since the elements of the DFT vector are approximately uncorrelated under stationarity on Euclidean spaces. Since isotropic models on spheres are uniquely characterized in terms of the spherical harmonic (SH) representation rather than a Fourier transform, it is natural to formulate a global test for isotropy based on the spherical harmonic coefficients. In our approach, we transform the data onto the spherical harmonic functions, which form a set of orthogonal basis functions on the sphere. We exploit the fact that the correlation between the coefficients is zero if the process is isotropic. Furthermore they are Gaussian if the random field we start with is Gaussian \citep{baldi2007some}. On the other hand, if the random field is not isotropic, this characterization will not hold. This prompts us to formulate our test based on the sample correlation structure among the spherical harmonic coefficients. This also ensures that the alternative considered in our test is very general since every anisotropic model has coefficients that are correlated in some way. We construct our test based on the largest eigenvalue of the sample correlation matrix, which increases as we move away from isotropy, giving us a right sided critical region for our test. The approach is computationally efficient for gridded data because fast Fourier transforms (FFT) can aid in the projection of the data onto spherical harmonics. The test can also be based on a manageable number of spherical harmonic coefficients, so no large dense matrices need
to be stored. We also show that the approximations employed in the test improve as the resolution of the data in space increases.

We apply the test on the near-surface air temperature projections for 2031-2035 obtained from the HadCM3 model. We do not expect the near-surface temperature data to be well-modeled by an isotropic model. However since we can build anisotropic models out of isotropic ones, we can apply the test on the isotropic component of anisotropic models to check anisotropic model assumptions. In this paper, we propose a sequence of anisotropic models for our temperature data, each more complex than the previous one and each having an isotropic process as model component. We apply the proposed test on the isotropic components of the models and consider the values of the test statistic to determine how well the models capture the anisotropy in the near-surface temperature fields.     

The rest of the paper is structured as follows. In Section 2, we discuss the HadCM3 dataset that we have used for this study. Section 3 illustrates our model and the test procedure. Section 4 presents a simulation study conducted to evaluate the performance of our test under various conditions. Section 5 presents our analysis
of the near surface air temperature data, where we include a thorough discussion of the nature of the anisotropies in the data. Section 6 gives conclusions and discussions. 
\bigskip

\section{Motivating Dataset}

The data set which motivated our idea is a part of the Coupled Model Intercomparison Project Phase 5 (CMIP5) archive. The CMIP 5 is a large multi-model ensemble project which has been used for the Intergovernmental Panel on Climate Change (IPCC) reports. The Hadley Centre Coupled Climate Model Version 3 (HadCM3) of the Met Office Hadley Centre (MOHC) is a coupled climate model that has been used considerably for various climate studies including climate prediction and climate modeling. HadCM3 was one of the significant models utilized as a part of the IPCC Third and Fourth Assessments, and furthermore adds to the Fifth Assessment. These models have a resolution of 2.5 degrees in latitude by 3.75 degrees in longitude, thereby producing a global grid of $73 \times 96$ grid cells. This is equivalent to a surface resolution of about $417 \text{ km } \times 278 \text{ km}$ at the Equator, reducing to $295 \text{ km } \times 278 \text{ km}$ at 45 degrees of latitude. These model simulations also consider a 360-day calender, where each month has 30 days.    

From the HadCM3 model outputs in CMIP 5, we consider the Representative Concentration Pathway 4.5 ('RCP4.5`) simulations of daily near-surface air temperature ('tas`) over the period of 2031-2035. The observations are in the Kelvin scale. As mentioned before, the temperature values are generated on a 73 $\times$ 96 latitude $\times$ longitude grid for 360 days per year, giving a total of approximately 12.6 million observations in the data set.

\bigskip

\section{Methodology}
\subsection{Spherical Harmonic Representation}
Let $Y_t(\theta, \phi), t \in {1, 2, \ldots, T}$ denote a Gaussian process (GP) on a sphere indexed by latitude $\theta \in [0, \pi]$ and longitude $\phi \in [0, 2\pi)$. The GP can be expressed in terms of spherical harmonic basis functions as suggested by \citet{jones1963stochastic}. Let $S_{l, m}(\theta, \phi)$ denote the Schmidt semi-normalized harmonics of degree $l$ and order $m$ on the surface of the sphere. Analytically, $S_{l, m}(\theta, \phi)$ can be defined as 
$$
S_{l,m}(\theta, \phi) = \twopartdef
{\sqrt{\frac{(l - m)!}{(l + m)!}} P_{l,m}(\mbox{cos} \theta) e^{im\phi}} {m \geq 0}
{(-1)^m S_{l,-m}^*(\theta, \phi)} {m < 0}
$$
where * denotes complex conjugation and $P_{l,m}(\mbox{cos} \theta)$ denotes the associated Legendre polynomial of degree $l = 0, 1, 2, \ldots$ and order $m = 0, 1, \ldots, l$, that is,
$$
P_{l,m}(x) = (-1)^m(1 - x^2)^{m/2}\frac{d^m}{dx^m}P_l(x),
$$
$$
P_l(x) = \frac{1}{2^l l!}\frac{d^l}{dx^l}(x^2 - 1)^l.    
$$

\noindent The spherical harmonics form a complete set of orthogonal basis functions on the sphere; in particular, 
$$
\int_{\theta = 0}^{\pi} \int_{\phi = 0}^{2\pi} S_{l, m}(\theta, \phi) S_{l', m'}(\theta, \phi)^{*}\mbox{sin}\theta d\phi d\theta = \frac{4\pi}{(2l + 1)}\delta_{l l'} \delta_{m m'}
$$
where $\delta_{ij} = \mathbbm{1}(i = j)$ is the Kronecker delta. As a result, processes defined on the sphere can be expressed in terms of expansions of the spherical harmonic functions. Here we consider
\begin{equation}\label{1}
Y_t(\theta, \phi) = \sum_{l = 0}^{\infty}\sum_{m = -l}^l a_{lmt}S_{l,m}(\theta, \phi)
\end{equation}
where $\lbrace a_{lmt}\rbrace$ is a triangular array (for each t), representing the set of complex-valued random spherical harmonic coefficients for which the sum in \eqref{1} converges in mean square.

The random variables $(a_{lmt})_{l,m}$ are uncorrelated and form a Gaussian family if and only if in addition to being Gaussian, $Y_t(\theta, \phi)$ is also isotropic \citep{baldi2007some}; also $$E\left[Re\text{ }(a_{lmt})\right] = 0 = E\left[Im \text{ } (a_{lmt})\right], m = 0, \ldots, l$$ and $Re\text{ }(a_{lmt})$ and $Im\text{ }(a_{lmt})$ are uncorrelated with  variance $$E\left[Re\text{ }(a_{lmt}) ^ 2\right] = E\left[Im \text{ } (a_{lmt}) ^ 2\right] = C_l/2$$ where $C_l$ is the power spectrum for degree $l$. Thus we have $\mbox{Var}(a_{lmt}) = E\left[ \left| a_{lmt} \right| ^ 2 \right] = E\left[Re\text{ }(a_{lmt}) ^ 2\right] + E\left[Im \text{ } (a_{lmt}) ^ 2 \right] = C_l$. Since the coefficients are uncorrelated across $l$ and $m$, the variance of $Y_t(\theta, \phi)$ is
\begin{align*}
\mbox{Var}(Y_t(\theta, \phi)) &= \sum_{l = 0}^{\infty} \sum_{m = -l}^l S_{l,m}(\theta, \phi) S_{l,m}^*(\theta, \phi) \mbox{Var} (a_{lmt})\\
&= \sum_{l = 0}^{\infty} \sum_{m = -l}^l S_{l,m}(\theta, \phi) S_{l,m}^*(\theta, \phi) C_l \\
&= \sum_{l = 0}^{\infty} C_l\sum_{m = -l}^l S_{l,m}(\theta, \phi) S_{l,m}^*(\theta, \phi)  \\
&= \sum_{l = 0}^{\infty} C_l,  \text{ by Uns\"old's Theorem \citep{unsold1927beitrage}. }  
\end{align*}

Our testing procedure relies on a transformation from the observations $Y_t(\theta, \phi)$ to the SH coefficients $a_{lmt}$. If $Y_t(\theta, \phi)$ were observed continuously over the sphere, then the spherical harmonic transform, given by 
$$
a_{lmt} = \int_{\theta = 0}^{\pi} \int_{\phi = 0}^{2\pi} Y_t(\theta, \phi) S_{l, m}(\theta, \phi) \mbox{sin} \theta d\phi d\theta 
$$
can be used to recover the coefficients $a_{lmt}$. However if we have data on a grid of size $s_1 \times s_2$, we cannot recover the coefficients exactly and so we estimate $a_{lmt}$ as the minimizer of 
\begin{equation}\label{2}
\sum_{i = 1}^{s_1 s_2} \left\lbrace Y_t(\theta_i, \phi_i) - \sum_{l = 0}^{l_{reg}} \sum_{m = -l}^l a_{lmt}S_{l,m}(\theta_i, \phi_i) \right\rbrace ^ 2 \bigtriangleup W_i
\end{equation}
where $\bigtriangleup W_i$ is the surface area of the $i^{th}$ quadrangle, relative to the surface area of the Earth and $l_{reg}$ is chosen so that $(l_{reg} + 1)^2 \leq s_1s_2$. For very large gridded datasets, the sums over longitudes and over $m$ can be computed efficiently with FFTs. This is a weighted least squares problem with the weights equal to the relative surface area of the quadrangles. 

Let $\bm{Y}_t$ denote the data vector for time $t$ at all spatial locations and $\bm{Y}$ be the $s_1 s_2 \times T$ matrix $[ \bm{Y}_1, \ldots, \bm{Y}_T ]$. Also let $\bm{S} = (S_{l, m})_{l, m}$ denote the matrix of the semi-normalized harmonics, truncated at degree $l_{reg}$ and $\bm{W}$ denote a diagonal matrix, with the weights $\bigtriangleup W_i$ on the diagonal. Then minimizing the sum with respect to $a_{lmt}$ gives the coefficient matrix as $\widehat{\bm{a}} = (\bm{S}'\bm{W}\bm{S})^{-1} \bm{S}'\bm{W} \bm{Y}$. We apply this transformation at each time point, and use $\widehat{\bm{a}}_{lm \bullet} = \left( \widehat{a}_{lm1}, \ldots \widehat{a}_{lmT}\right)$ to denote the SH coefficient corresponding to degree $l$ and order $m$, replicated over time, whereas we use $ \widehat{\bm{a}}_{\bullet t} $ to denote all the coefficients at time point $t$.
\bigskip

\subsection{Test Procedure}

Since we truncate the sum in \eqref{1} to represent the process, we work with a total of $n_{reg} = (l_{reg} + 1)^2$ spherical harmonics. We explore the selection of the truncation degree $l_{reg}$ in Section 4 based on the stability of the regression that converts $\bm{Y}$ to $\widehat{\bm{a}}$. Depending on the accuracy of the regression, we only use spherical harmonics up to degree $l_{corr} \leq l_{reg}$ and use $p = n_{corr} = (l_{corr}+1)^2$ coefficients in the test. The selection of $l_{corr}$ is also described in Section 4. Since the true coefficients $\bm{a}$ are uncorrelated under isotropy, our hypotheses about isotropy are equivalent to $$H_0: R = I_p \hspace{0.5cm} \text{  versus  } \hspace{0.5cm} H_1: R \neq I_p$$ where $R = \mbox{Corr}(\bm{a}_{\bullet t})$.

We construct the test statistic based on the eigenvalues of the sample correlation matrix of $\widehat{\bm{a}}_{\bullet 1}, \ldots \widehat{\bm{a}}_{\bullet T}$. Under the null hypothesis, the eigenvalues of the population correlation matrix will all be 1 and when we move away from the null, the largest sample eigenvalue will increase. This motivates us to form a test based on the largest sample eigenvalue. \citet{johnstone2001distribution} provides the distribution of the largest eigenvalue of the sample correlation matrix. For this purpose, let $\bm{w}_{lm \bullet}$ denote the standardized SH coefficient corresponding to degree $l$ and order $m$. Notationally, 
$$
\bm{w}_{lm \bullet} = \frac{\widehat{\bm{a}}_{lm \bullet}}{\|\widehat{\bm{a}}_{lm \bullet}\|}.
$$
Under $H_0$, the vectors $\bm{w}_{lm \bullet}$ are i.i.d. Now we multiply each standardized SH coefficient by an independent chi random variable in order to generate a standard Gaussian data matrix, denoted by $\tilde{\bm{a}}^{(p)} = (\tilde{\bm{a}}_{lm \bullet})_{l, m}$ where 
$$
\tilde{\bm{a}}_{lm \bullet} = r_{lm} \bm{w}_{lm \bullet}, \hspace{2cm} r_{lm}^2 \overset{indep}{\sim} \chi_T^2.
$$
The test statistic is $$\tilde{l}_1 = \frac{l_1(\tilde{\bm{C}}) - \mu_{Tp}}{\sigma_{Tp}}$$ where $l_1(\tilde{\bm{C}})$ is the largest sample eigenvalue of $\tilde{\bm{C}} = \tilde{\bm{a}}^{(p)'} \tilde{\bm{a}}^{(p)}$, $\mu_{Tp} = (\sqrt{T - 1} + \sqrt{p})^2$ and $ \sigma_{Tp} = (\sqrt{T - 1} + \sqrt{p})\left(\frac{1}{\sqrt{T - 1}} + \frac{1}{\sqrt{p}}\right)^{1/3}$. Under the null hypothesis, when $T$ and $p$ both increases such that $T/p \rightarrow \gamma \geq 1$,  $$\tilde{l}_1 \overset{d}{\rightarrow} W_1 \sim F_1 $$ 
where $F_1$ is the Tracy-Widom law of order 1 which is given by 
$$
F_1(s) = exp \left\lbrace \frac{1}{2} \int_s^\infty q(x) + (x - s)q^2(x)dx \right\rbrace , s \in \mathbb{R}
$$
where $q$ solves the nonlinear Painlev\'e II differential equation
$$
q''(x) = xq(x) + 2q^3(x),
$$
$$
q(x) \sim Ai(x) \text{ as x }\rightarrow +\infty
$$
and $Ai(x) = \frac{1}{\pi} \int_0^\infty cos\left(\frac{t^3}{3} + xt \right)dt$ denotes the Airy function.

The test is designed for $T \geq p$ but it applies equally well if $T < p$ are both large, simply by reversing the roles of $T$ and $p$ in the expressions for $\mu_{np}$ and $\sigma_{np}$ \citep{johnstone2001distribution}. The p-value for the test can be computed using the cumulative distribution table of the $TW_1$ distribution \citep{bejan2005}. Since the largest eigenvalue increases as we move away from isotropy, we have a right-tailed test. 

\section{Simulation Study}
\subsection{Stability of the SH estimates}

In this section we first justify that the weighted regression technique in \eqref{2} gives accurate coefficient estimates $\widehat{\bm{a}}$ when we truncate the sum in \eqref{1}. For this purpose, we choose $l_{sim} > l_{reg}$ and simulate $n_{sim} = (l_{sim} + 1)^2$ Gaussian complex-valued coefficients $\bm{a} (n_{sim} \times T)$ with variance $C_l$, independent over $T = 360$ time replicates and then apply \eqref{1} to get the spatial data, $Y_t(\theta,\phi)$ (forward transform). We evaluate how well we recover the $\bm{a}$ when we regress the data $\bm{Y}$ onto $\bm{S}$, the spherical harmonics truncated at $l_{reg}$ (back transform). 

For the variance of $Y_t(\theta, \phi)$ to exist, $C_l$ must be summable. To achieve these, we consider the variances
\begin{align*}
C_l = \frac{\sigma^2}{(\alpha^2 + l^2)^{\nu+1/2}},
\end{align*}
which gives rise to the Legendre-Mat\'ern covariance function \citep{guinness2016isotropic} given by
$$
\psi(\theta) = \sum_{l = 0}^{\infty} \frac{\sigma^2}{(\alpha^2 + l^2) ^ {\nu + 1/2}} P_l(\mbox{cos}\theta).
$$
Here $\sigma^2, \alpha, \nu > 0$ are the three parameters of the covariance function with $\sigma^2$ denoting the variance, $1/\alpha$ denoting the spatial range, and $\nu$, the smoothness. The form of the Legendre Mat\'ern is motivated by the Mat\'ern spectral density on $\mathbb{R}^d$, which is $(\alpha^2 + \omega^2) ^ {-\nu - 1/d}$. In particular, we take $\nu = 0.5$ and $\nu = 1$ for our simulation studies. This gives us $C_l$ of the order of $1/l^2$ and $1/l^3$ respectively. For our convenience we refer to the two spectra as $C_{l2}$ and $C_{l3}$ respectively. Note that the process obtained with $C_{l2} (\nu = 0.5)$ is not mean square differentiable. According to \citet{hitczenko2012some} this is similar to a process with exponential covariance. 

In order to ensure computational stability during regression (back transform), we choose the truncation degree of the SH, $l_{reg}$, based on the condition number of $\bm{S}'\bm{W}\bm{S}$, i.e., the ratio of its smallest eigenvalue to its largest. We choose the largest $l$ such that the condition number of $\bm{S}'\bm{W}\bm{S} > 0.001$. The regenerated coefficients can then be expressed as 
\begin{align*}
\widehat{\bm{a}}_{\bullet t} &= (\bm{S}'\bm{W} \bm{S})^{-1} \bm{S}' \bm{W} \bm{Y}_t\\
&= (\bm{S}'\bm{W} \bm{S})^{-1} \bm{S}' \bm{W} \bm{S} \bm{a}_{\bullet t}
\end{align*} 
Here $\widehat{\bm{a}}_{\bullet t}$ is of length $n_{reg}$ whereas $\bm{a}_{\bullet t}$ is of length $n_{sim}$, which is much larger than $n_{reg}$ since $l_{sim} > l_{reg}$.

The accuracy of the regression is summarized by the correlation between the unique real and imaginary parts of the true coefficients for each $l$ and the corresponding estimates, $$r_l = \frac{1}{2l + 1} \sum_{m = -l} ^ l Corr(\bm{a}_{lm \bullet}, \widehat{\bm{a}}_{lm \bullet})$$ where $$ Corr(\bm{a}_{lm \bullet}, \widehat{\bm{a}}_{lm \bullet}) = \frac{\sum_{t = 1}^{T} (a_{lmt} - \bar{a}_{lm})(\widehat{a}_{lmt} - \widehat{a}_{lm}^b)}{\sqrt{\sum_{i = 1}^{T} (a_{lmt} - \bar{a}_{lm})^2} \sqrt{\sum_{i = 1}^{T}(\widehat{a}_{lmt} - \widehat{a}_{lm}^b)^2}},$$ $\bar{a}_{lm} = \frac{1}{T} \sum_{t = 1}^T {a}_{lmt}$ and $\widehat{a}_{lm}^b = \frac{1}{T} \sum_{t = 1}^T \widehat{a}_{lmt}$ denote the means of $\bm{a}_{lm \bullet}$ and $\widehat{\bm{a}}_{lm \bullet}$ respectively. Figure 1 shows that the correlation between the true and estimated SH coefficients is a decreasing function of $l_{reg}$. This motivates us to choose $l_{corr}$ as the maximum degree of SH for which $r_l > 0.999$. This ensures that the weighted regression in \eqref{2} gives accurate SH coefficients as long as the degree of SH considered is less than or equal to $l_{corr}$. We also wish to study the effect of grid size on the performance of the weighted regression. With this in mind, we use three different grid sizes, namely $20 \times 50, 73 \times 96$ and $100 \times 200$ for our study. In all our numerical studies, $l_{sim}$ is chosen to be 150. In our data analysis, we have a grid of size $73 \times 96$. Our numerical study indicates that under spectrum $C_{l2}$ we can use $l_{corr}$ up to 30, which corresponds to constructing a test on $(30 + 1)^2 = 961$ unique coefficients. Figure 1 also shows that the regression performs better with increasing grid size and also with decreasing spectra. Table 1 illustrates how both the number of SH used for meaningful regression and accurate estimation of the coefficients grow with $l^2$.

\begin{figure}[h!]
\centering 
\begin{subfigure}{0.49\textwidth}
\centering
\includegraphics[width = \textwidth]{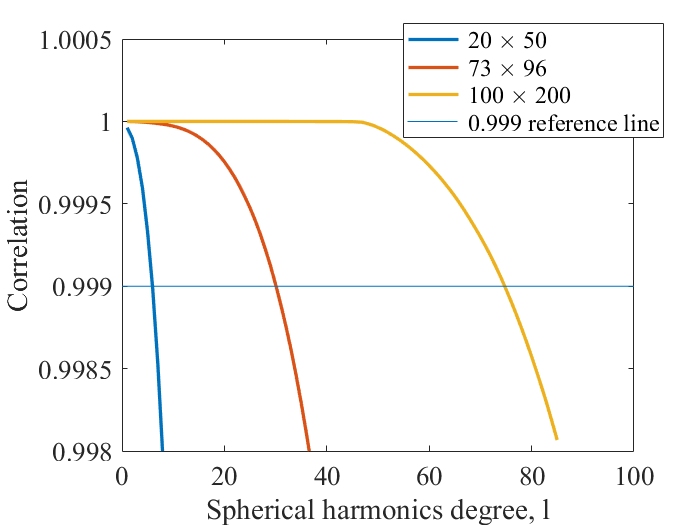}
\caption{Spectrum $C_{l2}$}
\end{subfigure}
\begin{subfigure}{0.49\textwidth}
\centering
\includegraphics[width = \textwidth]{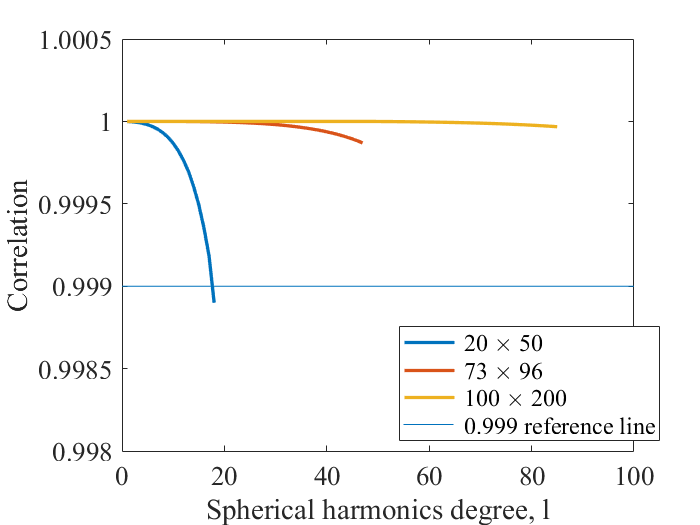}
\caption{Spectrum $C_{l3}$}
\end{subfigure}
\caption{Correlation between true and estimated coefficients for the three different grid sizes $20 \times 50$ (blue), $73 \times 96$ (red) and $100 \times 200$ (green) and for two different spectra (left versus right), as a function of the SH degree $l$. We consider the SH degree up to $l_{reg}$ in the weighted regression. For each grid size-spectra combination, we take $l_{corr}$ as that value of $l$ where the corresponding correlation curve intersects the $0.999$ reference line, which gives us $n_{corr} = (l_{corr} + 1)^2$ unique coefficients. }
\end{figure}

\begin{table}[H]
\caption{The maximum degree of SH that ensures computational stability during regression, $l_{reg}$, and the maximum degree of SH used to guarantee accurate estimation of the coefficients, $l_{corr}$, for the three different grid sizes and the two spectra. $n_{reg}$ and $n_{corr}$ give the number of SH functions used in each setting. $l_{sim}$ is chosen to be 150.}
\label{t1}
\centering

\begin{tabular}{ |c|c|c|c|c|c| } 
\hline 
Spectrum & Grid Size & $l_{reg}$ & $n_{reg}$ & $l_{corr}$ & $n_{corr}$ \\

\hline \hline
\multirow{3}{4em}{$C_{l2}$} & $20 \times 50$ & 18 & 361 & 6 & 49 \\

& $73 \times 96$ & 47 & 2304 & 30 & 961 \\
& $100 \times 200$ & 85 & 7396 & 74 & 5625 \\
\hline
\multirow{3}{4em}{$C_{l3}$} & $20 \times 50$ & 18 & 361 & 17 & 324 \\

& $73 \times 96$ & 47 & 2304 & 47 & 2304 \\

& $100 \times 200$ & 85 & 7396 & 85 & 7396 \\

\hline
\end{tabular}
\end{table}

In the next subsection, we assess the performance of our test. First, we calculate the Type I error of the test by generating time-independent coefficients under the null. Next, we consider temporal correlation among the coefficients. Finally, we compute the power of the test under anisotropic models. 

\subsection{Assessing the performance of the test}

\textbf{Type I error under no temporal correlation}

We simulate the coefficients as time-independent complex Gaussian, that is, 
\begin{align*}
Re\text{ }a_{lmt}, Im\text{ }a_{lmt} &\sim N(0, C_l/2), l = 1, \ldots, l_{max}, t = 1, \ldots, T = 360, \\
a_{00} &\sim N(0, 1.5)
\end{align*}
with $l_{sim}$ = 150 and for $C_{l2}$ and $C_{l3}$. We follow the test procedure as described in Section 3 for the three grid sizes $20 \times 50$, $73 \times 96$ and $100 \times 200$ with appropriate choices of $l_{reg}$ and $l_{corr}$ as described in Table 1. We perform the test at the 5\% significance level. The Type I error of the test is given by 
\begin{align*}
p &= Pr_{H_0}(TW(T, p) > T_{obs}) \\
  &= Pr_{H_0}\left( \frac{TW(T, p) - \mu_{Tp}}{\sigma_{Tp}} > \frac{T_{obs} - \mu{Tp}}{\sigma_{Tp}}\right) \\
  &= Pr_{H_0}\left( TW_1 > \frac{T_{obs} - \mu_{Tp}}{\sigma_{Tp}} \right)
\end{align*}

\noindent Table 2 shows the Type I error of the test for the three grid sizes and two different spectra based on 1000 simulation replications. The Type I error varies between 3\% and 7\% depending on the choice of $l$. For the recommended choice of $l_{corr}$ (the final entry in each column), the Type I error is very close to the nominal level in all cases. 

\begin{table}[t!]
\centering
\caption{Type I error (in \%) of the test for the three grid sizes, $20 \times 50$, $73 \times 96$ and $100 \times 200$ and the two spectra $C_{l2}$ and $C_{l3}$. We perform the test at the 5\% significance level. Here $l$ represents the SH degrees for which we perform the test. Note that for a particular setting, we only consider $l$'s which are less than or equal to the corresponding $l_{corr}$.}
\label{my-label}
\begin{tabular}{|c|c|c||c|c|c||c|c|c|}
\hline
\multicolumn{3}{|c||}{$20 \times 50$} & \multicolumn{3}{c||}{$73 \times 96$} & \multicolumn{3}{c|}{$100 \times 200$} \\ \hline \hline
 $l$     &   $C_{l2}$    &   $C_{l3}$    &    $l$   &   $C_{l2}$    &   $C_{l3}$    &   $l$    &   $C_{l2}$    &    $C_{l3}$   \\ \hline
  3 & 4.3 & 3.4 & 5  & 4.7 & 5.6 &   5  & 5.3 & 5.3 \\ 
  \hline
  4 & 5.2 & 3.6 & 10 & 4.8 & 3.7 &   10 & 5.4 & 5.4 \\ 
  \hline
  5 & 5.0 & 5.3 & 15 & 5.2 & 3.6 &   15 & 4.9 & 4.9 \\ 
  \hline
  8 &     & 4.3 & 20 & 4.7 & 3.7 &   25 & 3.6 & 3.6 \\ 
  \hline
 10 &     & 5.2 & 24 & 4.9 & 5.4 &   35 & 5.0 & 4.8 \\ 
 \hline
 15 &     & 5.8 & 27 & 4.8 & 5.6 &   45 & 4.9 & 4.9 \\ 
 \hline
    &     &     & 29 & 4.7 & 3.7 &   55 & 5.8 & 5.5 \\ 
    \hline
    &       &       &   35 &        & 4.5 &   65 & 4.9 & 4.7 \\ \hline
    &       &       &   40 &        & 4.3 &   70 & 4.7 & 4.8 \\ \hline
    &       &       &   45 &        & 4.4 &   74 & 5.6 & 5.5 \\ \hline
    &       &       &   47 &        & 5.2 &   80 &        & 4.2 \\ \hline
    &       &       &      &       &       &     85 &       & 5.0 \\ \hline
\end{tabular}
\end{table}

\medskip

\noindent \textbf{Type I error under temporal correlation}

Our test requires replications of the spatial process and for most applications the replications will be correlated in time. Based on our analysis of the climate temperature data in Section 5 and previous studies of space-time covariances \citep{stein2005} we expect the lower degree coefficients to have stronger temporal correlation than the higher degree coefficients. For our simulation study, we assume a simple AR(1) structure among the coefficients. For $t = 1, \ldots, T = 360$,
$$
a_{lmt} = \rho_l a_{lm(t - 1)} + e_{lmt}
$$
where the innovations $e_{lmt}$ are uncorrelated across l, m, and t, $e_{lmt} \sim CN(0, C_l)$ and $\rho_l = 0.9/\sqrt{l}, l = 1, 2, \ldots, l_{max}$ with $\rho_0 = 0.99,$ is the temporal correlation function which decays with the degree of the SH. The simulated data are transformed to real space using the forward-transform \eqref{1}. 

To illustrate the importance of addressing temporal dependence in spatio-temporal data, we perform the test directly on the coefficients obtained from back-transforming the data into SH coefficients. In such a scenario, the Type I error of the test is more than 99\% for each of the grid sizes and spectra, even when the underlying spatial covariance structure is isotropic. 

To account for temporal dependence in our test, we treat $\widehat{\bm{a}}_{lm \bullet}$, obtained from the back transform as a time series and estimate $\rho_l$ for every $(l, m)$ combination by regressing $\widehat{a}_{lm2}, \ldots, \widehat{a}_{lmT}$ on $\widehat{a}_{lm1}, \ldots, \widehat{a}_{lm(T - 1)}$. We then perform the test on the innovations at the 5\% significance level. Table 3 shows the Type I error of our test once we have accounted for temporal dependence. Once again we see that the test has the right size.  

\begin{table}[t!]
\centering
\caption{Type I error (in \%) of the test for the three grid sizes $20 \times 50$, $73 \times 96$ and $100 \times 200$ and the two different spectra, $C_{l2}$ and $C_{l3}$ after accounting for temporal correlation. We perform the test at 5\% significance level.}
\label{my-label}
\begin{tabular}{|c|c|c||c|c|c||c|c|c|}
\hline
\multicolumn{3}{|c||}{$20 \times 50$} & \multicolumn{3}{c||}{$73 \times 96$} & \multicolumn{3}{c|}{$100 \times 200$} \\ \hline \hline
 $l$      &   $C_{l2}$    &   $C_{l3}$    &    $l$   &   $C_{l2}$    &   $C_{l3}$    &   $l$    &   $C_{l2}$    &    $C_{l3}$   \\ \hline
  3 & 4.7 & 3.7 & 5  & 4.7 & 4.3 & 5  & 4.7 & 5.3      \\ \hline
  4 & 4.3 & 4.1 & 10 & 4.6 & 4.5 & 10 & 4.4 & 5.1      \\ \hline
  5 & 4.4 & 6.0 & 15 & 4.8 & 4.4 & 15 & 4.8 & 5.1      \\ \hline
  8 &     & 5.0 & 20 & 4.9 & 4.8 & 25 & 4.6 & 5.1      \\ \hline
  10 &    & 5.2 & 24 & 5.0 & 5.1 & 35 & 5.2 & 4.9      \\ \hline
  15 &    & 5.9 & 27 & 5.3 & 5.0 & 45 & 4.9 & 5.4      \\ \hline
     &    &     & 29 & 4.5 & 5.4 & 55 & 4.4 & 5.0      \\ \hline
     &    &     & 35 &     & 5.7 & 65 & 4.2 & 4.5      \\ \hline
     &    &     & 40 &     & 4.3 & 70 & 4.1 & 4.8      \\ \hline
     &    &     & 45 &     & 5.7 & 74 & 4.9 & 5.0      \\ \hline
     &    &     & 47 &     & 5.3 & 80 &     & 4.3      \\ \hline
     &    &     &    &     &     & 85 &     & 4.3      \\ \hline
\end{tabular}
\end{table}

\medskip

\noindent \textbf{Power computations}

In this section we study the power of our test for anisotropic data. We consider two simple anisotropic scenarios. In the first scenario we introduce anisotropy by incorporating a correlation among the coefficients. In particular, we generate the $a_{lmt}$'s as complex Gaussian with variance $C_{l2}$ and 
$$
\mbox{Corr}(a_{lmt}, a_{lm't}) = \twopartdef
{1,}{m = m'}
{\psi,}{m \neq m', 0 \leq \psi \leq 1}.
$$

\noindent Figure 2 plots the power by  $\psi$ for the three grid sizes $20 \times 50$, $73 \times 96$ and $100 \times 200$. All results are based on 1000 simulation replications and $T = 360$ independent time replications for each simulation replication. For each of the 1000 datasets we conduct the test with suitable $l_{reg}$ as mentioned in Table 1 and a few suitable $l$'s as listed in Table 2.

\begin{figure}[h!]
  \centering
  \includegraphics[width = \textwidth]{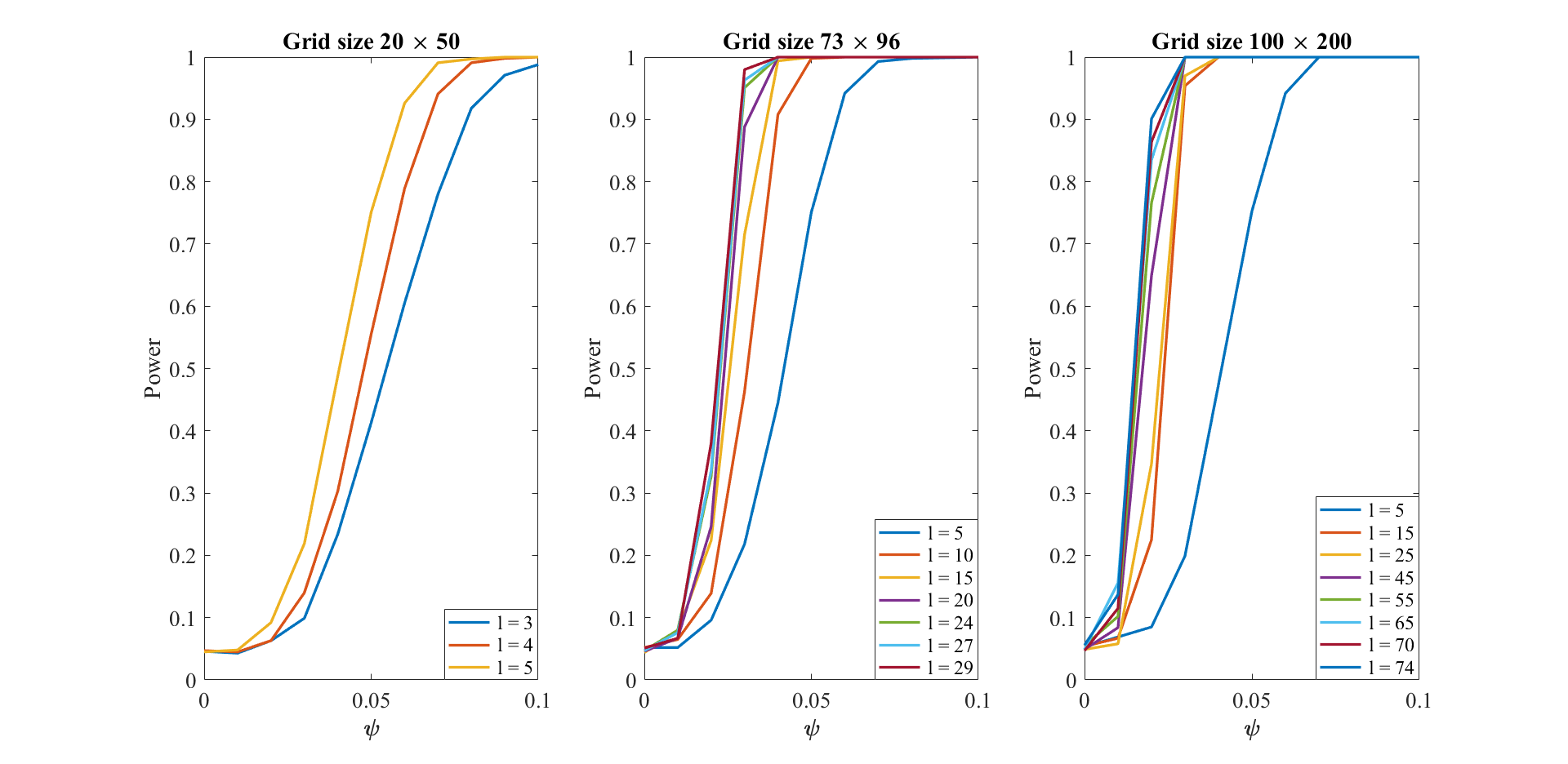}
  \caption{Empirical power functions of our test (as a function of $\psi$) corresponding to Scenario 1 for the three grid sizes and different degrees of SH, $l$.}
\end{figure}

\noindent Figure 2 shows that the test is very powerful in detecting even the slightest departures from isotropy. Even if the true correlation between two SH coefficients in the same degree is as small as $0.05$, the test can almost always detect that the process has deviated from isotropy for a reasonable grid size and with a relatively small degree of the spherical harmonics. We see that the power increases with the degree of SH functions used in our analysis. The power also increases as the data points on the sphere becomes more dense. 

Another way to introduce anisotropy directly in the fields is to assume that the covariance structures over different parts of the globe are different. A simple way to do it is to consider different covariances over land and water. In particular, we define 
$$
g_l(s) = \twopartdef
{1, }{\text{if } s \in land}
{1/l^\epsilon, }{\text{if } s \in ocean}
$$
where $\epsilon \geq 0$; $\epsilon = 0$ gives back the case of isotropy. The fields are then generated as $$Y_{\mbox{aniso}; t}(s) = \sum_{l = 0}^{l_{sim}} \sum_{m = -l}^l g_l(s) a_{lmt}S_{l, m}(\theta, \phi).$$ where $a_{lmt}, t = 1, \ldots, T = 360$ are simulated as complex Gaussian with variance $C_{l2}$ and independent over time. Since $\epsilon > 0$, $g_l(s)$ has the effect of reducing variance of high frequency coefficients, resulting in smoothing processes over the ocean. Figure 3 plots the empirical power functions of our test as a function of $\epsilon$ with other settings the same as for Figure 2. Once again we consider 1000 simulations replications for estimating the power function. Once again, the test is very powerful even for very minor departures from isotropy with $\epsilon < 0.1$. We also see that the power of the test increases with the degree of the SH used in our analysis and the grid size.  

\begin{figure}[h!]
  \centering
  \includegraphics[width = \textwidth]{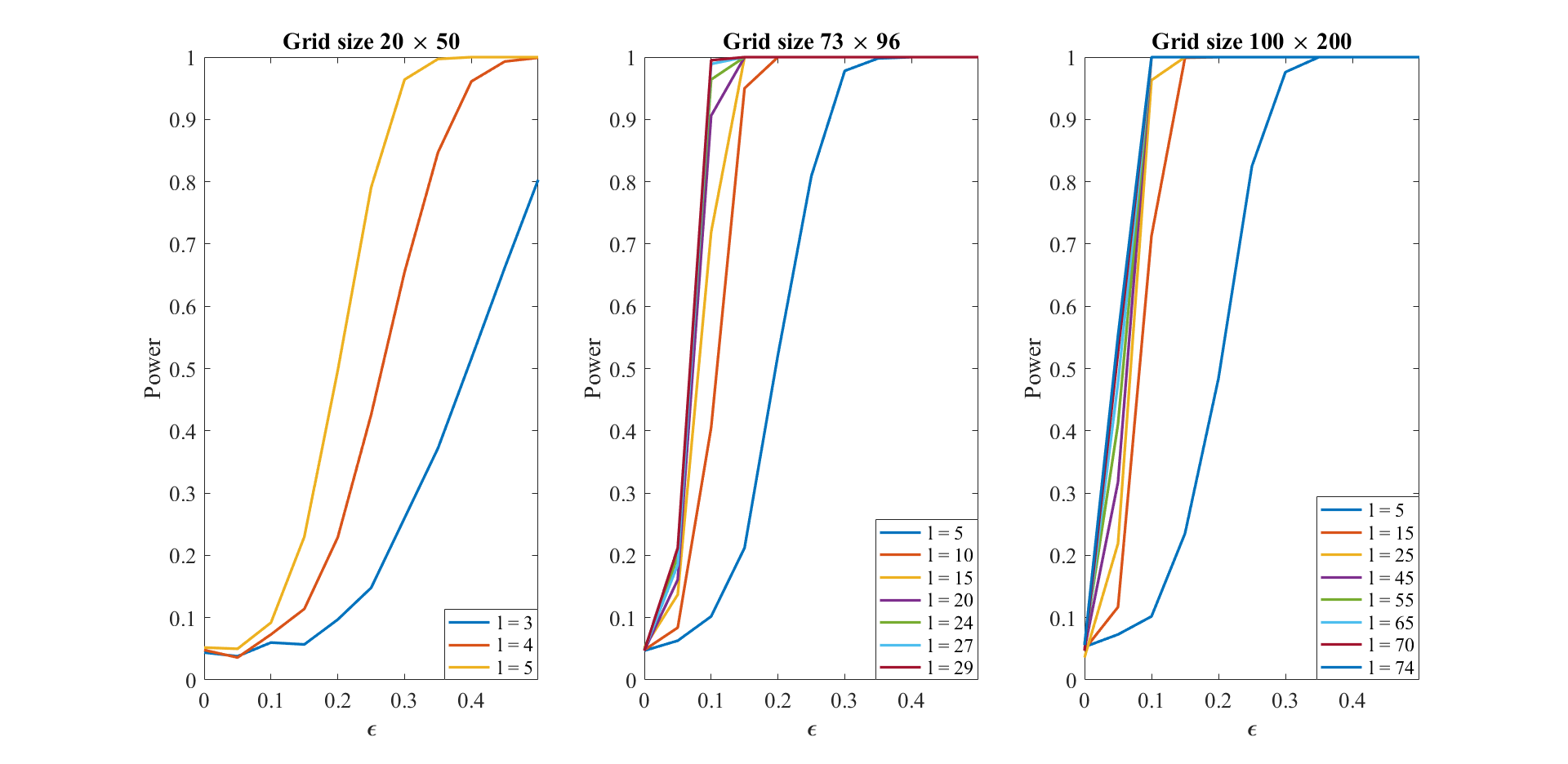}
  \caption{Empirical power functions of our test (as a function of $\epsilon$) corresponding to Scenario 2 for the three grid sizes and different degrees of SH, $l$. }
\end{figure}

\bigskip

\section{Application to HadCM3 Model Output data}

We apply our method to the near-surface air temperature data obtained as an output of the HadCM3 climate model. We work with daily air temperature data from 2031 - 2035 projected on a $73 \times 96$ grid in latitude and longitude. Each month in the data has 30 days. Thus we have 5 years worth of data with 360 time points corresponding to each year, resulting in $T = 1800$ time points. While we do not believe that the temperature fields are isotropic, we use the test to estimate the goodness of fit of models which seek to remove the anisotropies in the fields. We consider a few anisotropic models based on isotropic processes and we perform the test on the isotropic component of each model. In each of the models, $Y_t(\theta, \phi), t = 1, \ldots, T$ denote the near-surface air temperature at location $(\theta, \phi), \theta \in [0, \pi], \phi \in [0, 2\pi)$. We consider three models of increasing complexity, and each model $M_i$ can be written in the form

$$
Y_t(\theta, \phi) = m_t^{(i)}(\theta, \phi) + e_t^{(i)}(\theta, \phi) 
$$
$$
\frac{1}{k^{(i)}(\theta, \phi)}e_t^{(i)}(\theta, \phi) = \sum_l \sum_m a_{lmt}S_{l,m}(\theta, \phi)
$$
$$
a_{lmt} = \rho_{lm} a_{lm (t - 1)} + \epsilon_{lmt}
$$
where $\epsilon_{lmt} \sim N \left( 0, \sigma^2_{lm} \right)$. Table 4 describes the form of $m_t^{(i)}(\theta, \phi)$ and $k^{(i)}(\theta, \phi)$ for each of the models considered.

\begin{table}[t!]
\centering
\caption{Description of the three models considered. Here, $b_0, b_1, b_2 \in \mathbb{R}$ and $\sigma^2(\theta, \phi)$ denotes the spatially varying variance.}
\begin{tabular}{|c|c|}
\hline
Model Name    &  Model Description\\
\hline \hline         
\multirow{2}{*}{$M_1$} & $ m_t^{(1)}(\theta, \phi) = b_0(\theta, \phi) $\\
                  &  $ k^{(1)}(\theta, \phi) = 1 $\\ \hline
\multirow{2}{*}{$M_2$} & $ m_t^{(2)}(\theta, \phi) = b_0(\theta, \phi) + b_1(\theta, \phi) \mbox{sin}\left( 2\pi\frac{t}{360} \right) + b_2(\theta, \phi) \mbox{cos}\left( 2\pi\frac{t}{360} \right) $ \\
                  & $ k^{(2)}(\theta, \phi) = 1 $ \\ \hline
\multirow{2}{*}{$M_3$} & $ m_t^{(3)}(\theta, \phi) = b_0(\theta, \phi) + b_1(\theta, \phi) \mbox{sin}\left( 2\pi\frac{t}{360} \right) + b_2(\theta, \phi) \mbox{cos}\left( 2\pi\frac{t}{360} \right) $ \\
                  & $ k^{(3)}(\theta, \phi) = \sigma(\theta, \phi) $ \\
\hline               
\end{tabular}
\end{table}

\noindent The first model $M_1$ has only a spatially varying mean, $M_2$ has the pixel-wise seasonal variation in its mean structure along with the spatially varying mean. $M_3$ additionally takes into account the spatially varying variance. In Section 4.2 we have already discussed how temporal correlation in the data, if not accounted for, can produce misleading results when it comes to checking if a process is indeed isotropic. Thus in each one of our models, we model the temporal dependencies in the SH coefficients as AR(1), assuming that the AR coefficients and the innovation variance vary with each $(l, m)$ combination. For each of the models, the spatially varying mean, $a_0$ is estimated by the pixel-wise mean temperature, $$\bar{Y}(\theta, \phi) = \frac{1}{1800} \sum_{t = 1}^{1800} Y_t(\theta, \phi)$$ which is illustrated in Figure 4. The other parameters in $M_2$ and $M_3$, namely $a_1, a_2, \rho_{lm}$ and $\sigma^2_{lm}$ have been estimated at each pixel by regressing the last $T - 1$ SH coefficients on the first $T - 1$. We use the SH degree of $l = 25$ and work with $(l + 1)^2 = 26^2 = 676$ SH coefficients which means that the sample correlation matrix of the coefficients is $676 \times 676$ for each of the three models. The test when applied on the isotropic components of the models yields the test statistic values 958.83, 716.53 and 328.63 for $M_1$, $M_2$ and $M_3$ respectively. This shows that as the model complexity increases, the models do a better job at explaining the anisotropy in the temperature fields. However, even for the most complex model, we still get a very strong rejection of the isotropic component. The AR(1) coefficient estimates and the estimated innovation variance corresponding to the different degrees of SH for $M_3$ are shown in Figure 5(a).

\begin{figure}[h!]
	\centering
    \includegraphics[width = \textwidth]{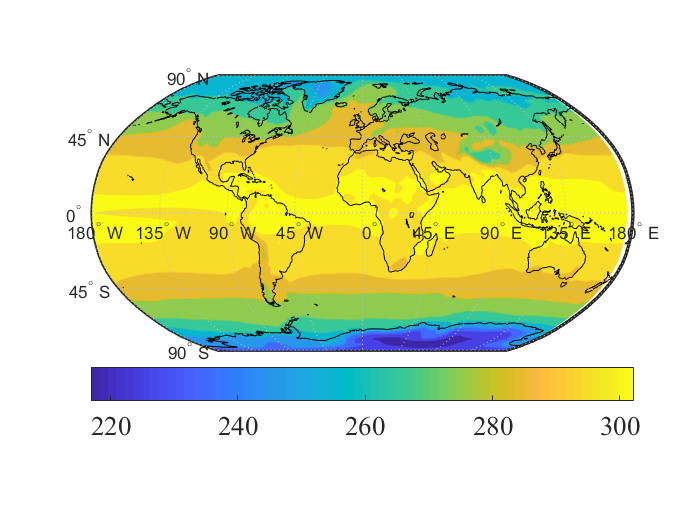}
    \caption{Pixel-wise mean air temperature (in Kelvin) based on the entire five years worth of model-output data}
\end{figure}

\begin{figure}[h!]
\centering
% \begin{adjustbox}{width=1.2\textwidth,center=\textwidth} 
\begin{subfigure}{\textwidth}
\centering
\includegraphics[width = \textwidth]{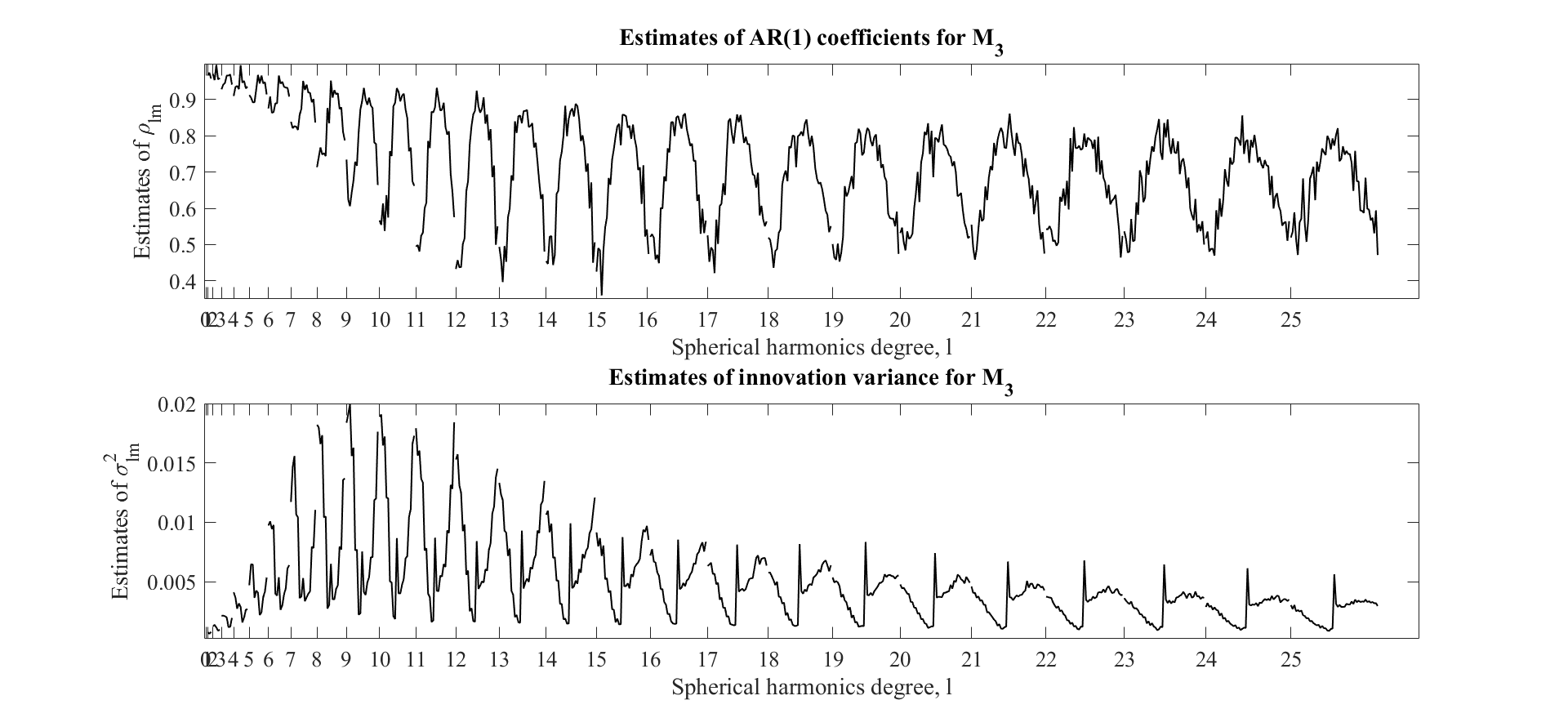}
\caption{Parameter estimates for $M_3$}
\end{subfigure}

\begin{subfigure}{0.65\textwidth}
\centering
\includegraphics[width = \textwidth]{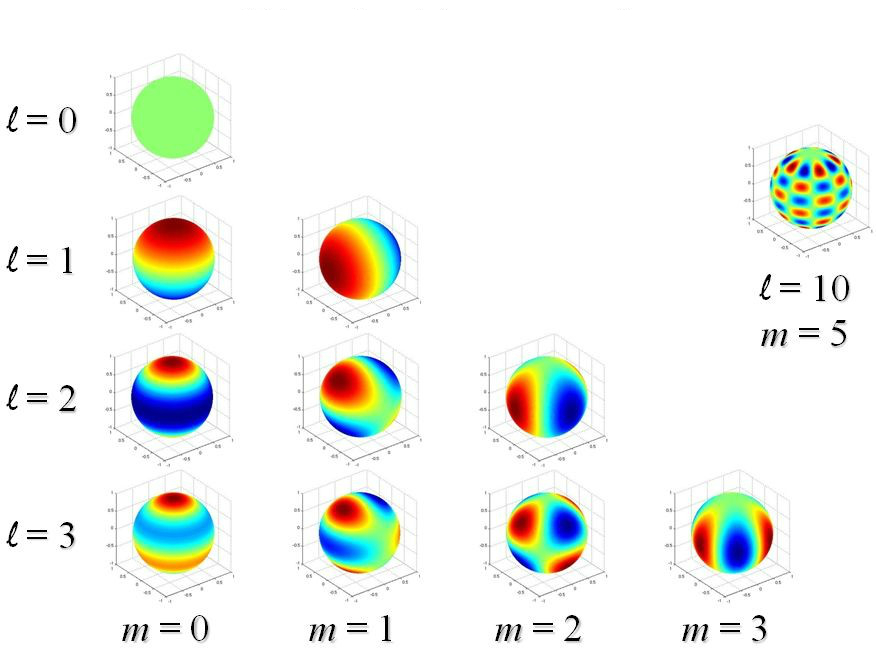}
\caption{Spherical harmonic functions}
\end{subfigure}
\caption{(a) Estimates of $\rho_{lm}$ and $\sigma_{lm}^2$ corresponding to $m = -l, \ldots, -1, 0, 1, \ldots, l$ under each $l$ for $M_3$. (b) Spherical harmonic functions for $m = 0, \ldots, l$ for $l = 0, \ldots, 3$. The spherical harmonics for negative $m$ can be depicted by rotating the positive order ones along the z-axis by $90\degree/m$. The checkerboard pattern has been shown for $l = 10, m = 5$. }
% \end{adjustbox}
\end{figure}

Figure 5(b) enables us to visualize the SH functions more clearly. When the spherical harmonic order $m$ is zero, the SH functions do not vary with longitude. Also with the increase in $\lvert m \rvert$, the SH functions start to have more spherical harmonics along the longitudinal axis and converge to 0 at the poles at a faster rate creating checkerboard pattern on the sphere until $l = \lvert m \rvert$ has all the harmonics along the longitude. Figure 5(a) shows that for each degree, the $m = 0$ coefficient is the most correlated in time and the dependence goes down with the increase in $\lvert m \rvert$. One can say that the spectral representation is analogous to the two-dimensional Fourier transform where each combination of the pair $(m, n)$ corresponds to a two-dimensional frequency. Thus, based on Figure 5(a) we can say that the low frequency coefficients are very highly correlated compared to the high frequency ones. Figure 5(a) also shows that within each $l$, the temporal correlation is maximum for $m = 0$ and it decreases as $\lvert m \rvert$ increases. Since the spherical harmonics are aligned along the latitudinal direction for $m = 0$ and start to get aligned along the longitudes as $\lvert m \rvert$ increases, we can say that the temperature process is more correlated along the direction of the latitudes as compared to the direction of the longitudes. Figure 5(a) also shows the power spectrum of the spectral representation for Model $M_3$. It shows the strength of each frequency signal and tells us that lower frequencies are, in general, more important than the higher ones. In particular, the spherical harmonics between degrees 7 to 12 seem to be most meaningful in explaining the temperature process. The spectral densities for models $M_1$ and $M_2$ have dominant peaks at low frequencies, thereby overshadowing all other peaks. This is due to low frequency variation not
captured by the spatially varying variance.

The innovations from $M_3$ are still anisotropic and we point out a few locations attributable to the anisotropy in the process. In order to get an estimate of the anisotropic covariance of the innovation process, we use the covariance of the innovation coefficients. If $\bm{a}_{innov}$ denote the innovation coefficients, then the estimate of the covariance in the data illustrating the remaining anisotropy is given by $Cov_{ani} = \bm{S} Cov(\bm{a}_{innov}) \bm{S}'$. On the contrary, if the process were isotropic at this stage, the covariance matrix of the innovation coefficients would be diagonal. So in order to get an estimate of the hypothetical isotropic covariance, we shrink all the off-diagonal elements of $Cov(\bm{a}_{innov})$ to zero. The locations with large deviation between the absolute values of the estimated covariances can be thought to be the top sites contributing to the anisotropy of the near-surface air temperature fields on the Earth. Figure 6 shows these locations along with the anisotropic and isotropic spatial covariances. 

\begin{figure}[h!]
	\centering
    \includegraphics[width = 0.95\textwidth]{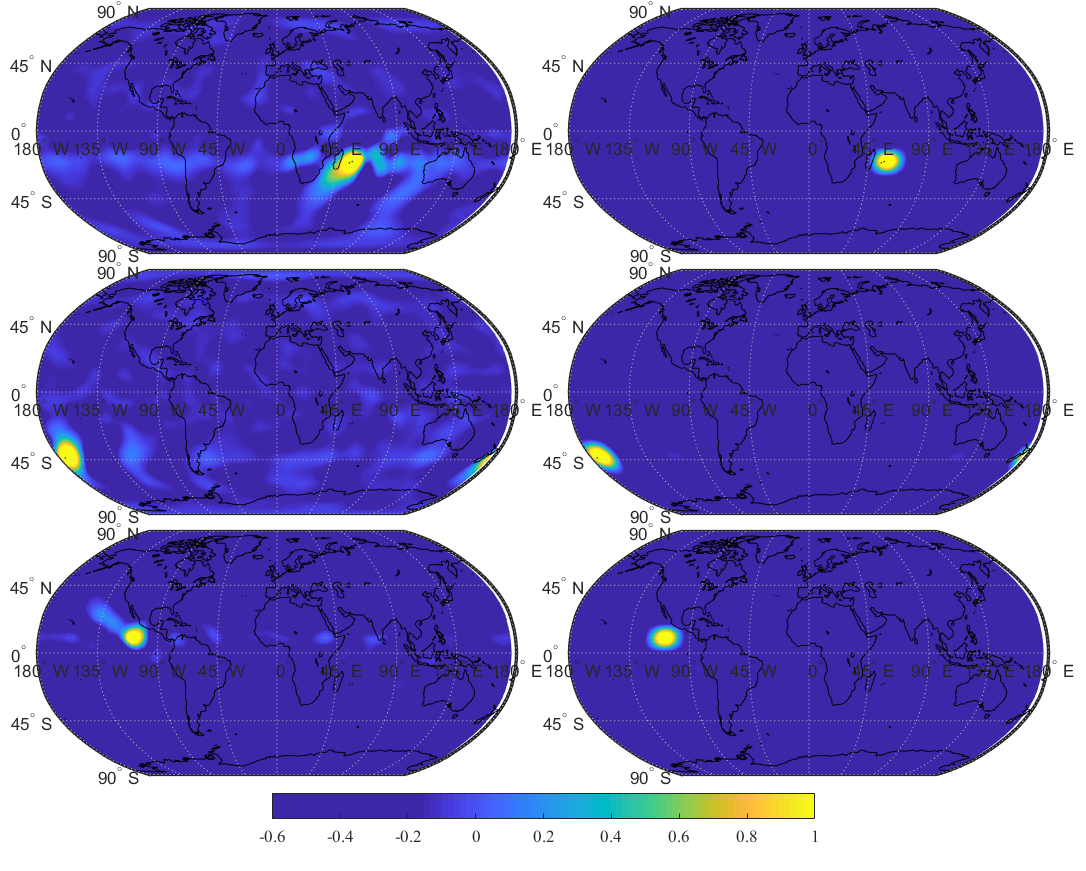}
  \caption{Estimates of the anisotropic (left panel) and (hypothetical) isotropic (right panel) correlation functions at three locations around the globe, namely northeast of Mauritius (first row), around the 45\degree S latitude and the International Date line (second row) and North Pacific Ocean, off the coast of Mexico (third row). }
\end{figure}

The first location chosen is just to the north-east of the islands of R\'eunion and Mauritius located to the east of Madagascar in the western Indian Ocean, the covariance structures of which are shown in Figure 6, row 1. The near-surface air temperature anomalies in this region can be associated with outgoing longwave radiation (OLR) anomalies over the west Pacific Ocean \citep{misra2004teleconnection}. This can also be combined with the possibility that rainfall anomalies over eastern South Africa can potentially affect temperatures in the western and south-western Indian Ocean \citep{reason1999relationships}. Row 2 of Figure 6 corresponds to our second location which is in the south Pacific Ocean just above the 45\degree S latitude and slightly to the right of the International Date line. This can be linked to low-frequency variation in the atmospheric circulation over the Southern Hemisphere extratropics \citep{carleton2003atmospheric}. This coincides with the Southern Oscillation, which is characterized by the barometric difference between Darwin and Tahiti. Fluctuations in this difference cause temperature anomalies in parts of western Pacific and hence might lead to large anisotropies in the temperature covariance. The third location (Figure 6, row 3) is in the North Pacific Ocean, off the coast of Mexico. This location is at the junction of the Pacific/North American (PNA) teleconnection pattern prevalent over the central North Pacific and the equatorial Pacific Ocean which is the El Ni\~no zone. This should account for the temperature anomalies in this area which throws the covariance structure in the temperature fields away from isotropy. 

\bigskip

\section{Discussions and Conclusions}

With the availability of large-scale global climate data, it is necessary to develop spatiotemporal models on a sphere that will explain the underlying spatial process and help make accurate predictions. It might be convenient to assume that the covariance structure on the globe is isotropic. However in most real-life  applications, this assumption does not hold. In this paper, we have proposed a method to determine the aptness of this simplifying assumption.  

We assume that a particular meteorological variable is distributed as a GP on a sphere and we express the process as a linear combination of the spherical harmonic functions which form a complete set of orthogonal basis functions on the sphere. Under the further assumption of isotropy, the spherical harmonic coefficients are uncorrelated, Gaussian \citep{baldi2007some}. We use this characterization of the coefficients to set up a test for isotropy based on the sample correlation among the coefficients. The test statistic, based on \cite{johnstone2001distribution} is given in Section 3.2. We provide conditions to ensure computational stability and accuracy during regression in Section 4.1.

In Section 4.2 we perform an extensive simulation study to evaluate the performance of our test. We look at the Type I error for three grid sizes and two different spectra under both time independence and considering a simple AR(1) dependence in time. The grid sizes we consider are similar to the data-resolutions that we generally observe in real-life applications. Our simulation results show that the test has the right Type I error for all the three grid sizes and under the different conditions. We also consider the power of our test under two anisotropic models. In the first case, we assume a correlation structure among the spherical harmonic coefficients which gives us a class of anisotropic models on the sphere. In the next case, we consider that the covariance across the globe is different on land and water. Figures 2 and 3 show that the test is able to detect slight deviations from isotropy in both the scenarios. It can also be seen that the power of the test increases as the resolution of the grids become finer. Also the power increases with the number of spherical harmonics coefficients used to compute the sample correlation matrix, which in turn depends on the maximum degree of the SH considered. 

We show how the test is sensitive to temporal correlation, and we have provided a modeling framework for addressing temporal correlation that gives accurate Type I error rates. This has been demonstrated in Section 4.2 where we consider a decaying temporal correlation among the coefficients and perform the test before and after modeling the temporal dependencies. It must be very evident that most spatio-temporal processes are not isotropic and our method provides a way to objectively perform a test to help arrive at that conclusion. Also one can easily arrive at the possible locations in the data attributing to the anisotropy using our method. As seen in our data analysis, even for the most complex model considered, the test for isotropy gets rejected. This highlights the need for developing better anisotropic models which will better capture the global anisotropic covariance structures of spatiotemporal processes. 

\newpage

\noindent {\large\bf Acknowledgements}

\medskip 

We would like to thank Dr. Dorit Hammerling for bringing the CMIP 5 data archive to our attention and also for her insightful comments during the course of this work. We also acknowledge the World Climate Research Programme's Working Group on Coupled modeling, which is responsible for CMIP, and we thank the MOHC for producing and making available their model output. This material is based upon work supported by the National Science Foundation under Grant No. 1613219. Additionally, Reich and Guinness were partially supported by National Institutes of Health Grant R01ES027892.

%\iffalse
\bibhang=1.7pc
\bibsep=2pt
\fontsize{9}{14pt plus.8pt minus .6pt}\selectfont
\renewcommand\bibname{\large \bf References}
%\begin{thebibliography}{11}
\expandafter\ifx\csname
natexlab\endcsname\relax\def\natexlab#1{#1}\fi
\expandafter\ifx\csname url\endcsname\relax
  \def\url#1{\texttt{#1}}\fi
\expandafter\ifx\csname urlprefix\endcsname\relax\def\urlprefix{URL}\fi
%\fi

\end{document}